\documentclass[twocolumn,aps,prd,preprintnumbers,superscriptaddress,nofootinbib,amsmath,amssymb,floats,floatfix,showkeys,notitlepage,showpacs]{revtex4-1}
\usepackage{orcidlink}
\usepackage{comment}
\usepackage{lipsum}
\usepackage{graphicx}
\usepackage{subfigure}
\usepackage{palatino}
\usepackage{hyperref}
\hypersetup{colorlinks=true,linkcolor=blue,urlcolor=blue,citecolor=blue}
\usepackage[utf8]{inputenc}
\usepackage{latexsym}
\usepackage{amsfonts}
\usepackage{dcolumn}
\usepackage{bm}
\usepackage{bigints}
\usepackage{array,tabularx,multirow,booktabs}
\usepackage[tracking=true]{microtype}
\usepackage{multirow}
\SetTracking{}{500}
\SetTracking{encoding={*}, shape=sc}{40}
\UseRawInputEncoding 
\allowdisplaybreaks

 \newcommand{\bq}{\begin{equation}}
 \newcommand{\eq}{\end{equation}}
 \newcommand{\bqn}{\begin{equation}}
 \newcommand{\eqn}{\end{equation}}

\newcommand{\be}{\nopagebreak[3]\begin{equation}}
\newcommand{\ee}{\end{equation}}

\newcommand{\ba}{\nopagebreak[3]\begin{equation}}
\newcommand{\ea}{\end{equation}}

\NewDocumentCommand{\evalat}{sO{\big}mm}{%
  \IfBooleanTF{#1}
   {\mleft. #3 \mright|_{#4}}
   {#3#2|_{#4}}%
}

\begin{document} \sloppy
	\newcommand \nn{\nonumber}
	\newcommand \fc{\frac}
	\newcommand \lt{\left}
	\newcommand \rt{\right}
	\newcommand \pd{\partial}
	\newcommand \e{\text{e}}
	\newcommand \hmn{h_{\mu\nu}}
	
	\newcommand{\PC}[1]{\ensuremath{\left(#1\right)}} 
	\newcommand{\PX}[1]{\ensuremath{\left\lbrace#1\right\rbrace}} 
	\newcommand{\BR}[1]{\ensuremath{\left\langle#1\right\vert}} 
	\newcommand{\KT}[1]{\ensuremath{\left\vert#1\right\rangle}} 
	\newcommand{\MD}[1]{\ensuremath{\left\vert#1\right\vert}} 

\title{Probing Weak-Force Corrections to Black Hole Geometry via Long Range Potentials: Feinberg-Sucher and Ferrer-Nowakowski Potentials}

\author{Ali \"Ovg\"un \orcidlink{0000-0002-9889-342X}}
\email{ali.ovgun@emu.edu.tr}
\affiliation{Physics Department, Eastern Mediterranean University, Famagusta, 99628 North
Cyprus via Mersin 10, Turkiye.}

\begin{abstract}
We consider corrections to the Schwarzschild black hole metric arising from exotic long-range forces within quantum field theory frameworks. Specifically, we analyze two models: the Feinberg-Sucher potential for massless neutrinos and Ferrer-Nowakowski potentials for boson-mediated interactions at finite temperatures, yielding metric corrections with $r^{-5}$ and $r^{-3}$ dependencies. Using analytic expansions around the Schwarzschild photon sphere, we find that attractive potential corrections enhance gravitational lensing, enlarging the photon sphere and shadow radius, while repulsive potential corrections induce gravitational screening, reducing these observables. Our results clearly illustrate how different quantum-derived corrections can produce measurable deviations from standard Schwarzschild predictions, providing robust theoretical benchmarks for future astrophysical observations.
\end{abstract}


\date{\today}
\maketitle

\section{Introduction}

Black holes have long served as fundamental laboratories for testing our understanding of gravity, quantum mechanics, and cosmology. Recently, the observational capability to image black hole shadows, initiated by the groundbreaking observations from the Event Horizon Telescope (EHT) \cite{EventHorizonTelescope:2019dse,EventHorizonTelescope:2022xqj}, has significantly expanded opportunities for probing gravity in strong-field regimes \cite{Vagnozzi:2022moj}. Innovative methodologies and theoretical frameworks continue to emerge, aiming to interpret observational data accurately and explore potential signatures of new physics, such as exotic long-range forces \textcolor{black}{arising from the Standard Model and beyond.}
The direct observation of black hole shadows has become a pivotal tool in testing gravitational theories and probing the nature of compact astrophysical objects. 

The landmark imaging of the supermassive black holes in M87* and Sgr A*~\cite{Falcke:1999pj,Chen:2022nbb,Zakharov:2018syb,Zakharov:2014lqa,Bambi:2019tjh,Gralla:2019xty} has significantly accelerated theoretical investigations, placing new constraints on potential deviations from the Kerr hypothesis~\cite{Johannsen:2010ru,Cardoso:2016ryw,Kuang:2022ojj}. Black hole shadows serve as observable signatures of photon spheres, sensitive to strong gravitational lensing effects and the geometry of spacetime near the event horizon~\cite{Cunha:2018acu,Cunha:2016wzk,Perlick:2021aok,Wei:2018xks}. Numerous theoretical and numerical studies have explored black hole shadows in diverse gravitational theories, including Einstein-Gauss-Bonnet gravity~\cite{Konoplya:2020bxa,Cunha:2017eoe}, non-linear electrodynamics~\cite{Okyay:2021nnh,Jafarzade:2020ova}, braneworld scenarios~\cite{Amarilla:2011fx}, and modified gravity models~\cite{Hennigar:2018hza}. The recent analyses have also considered plasma effects, scalar hair, and surrounding dark matter distributions on shadow observations~\cite{Shaikh:2018lcc,Konoplya:2019sns,Khodadi:2020jij,Afrin:2021imp,Zeng:2020dco}. In this context, accurate computations of photon spheres and black hole shadows not only test gravitational theories but may also reveal the presence of extra dimensions or new physics beyond the standard model~\cite{Vagnozzi:2019apd,Allahyari:2019jqz,Wei:2013kza,Belhaj:2020okh}. 
The formation mechanisms of regular black holes from baryonic matter have recently been explored, emphasizing their astrophysical relevance and stability under realistic conditions~\cite{Vertogradov:2025snh}. Additionally, analytical methods have been developed to characterize the shadow of dynamical black holes, providing novel insights into how evolving black hole parameters affect observable signatures~\cite{Vertogradov:2024eim}. Shadow of the exact regular black hole solutions with de Sitter cores have been investigated using Hagedorn fluid models, demonstrating new solutions that avoid singularities and maintain regularity at the core~\cite{Vertogradov:2024seh}. Furthermore, recent studies have developed general analytical approaches for analyzing the impact of geometric deformations on the photon sphere and shadow radius in both stationary, axisymmetric spacetimes~\cite{Vertogradov:2024jzj}, and spherically symmetric spacetimes~\cite{Vertogradov:2024qpf}. Such works significantly advance our understanding of spacetime geometry near black holes. Generalized frameworks have also been established for calculating shadow radii and photon spheres in asymptotically flat spacetimes, highlighting the influence of mass-dependent variations in these systems~\cite{Vertogradov:2024dpa}. Complementing these theoretical advancements, recent analyses show how primary hair and plasma distributions influence the intensity distribution within black hole shadows, offering potentially observable features distinct from classical predictions~\cite{Vertogradov:2024fva}. Moreover, dynamical photon spheres in charged black holes and naked singularities have been examined, revealing conditions under which these features evolve significantly, providing critical tests for theories of gravity and black hole physics~\cite{Heydarzade:2023gmd}. The effects of noncommutative geometry introduce smeared mass distributions replacing conventional singularities in Schwarzschild black holes~\cite{Nicolini:2005vd}, while Yukawa-type potentials offer modifications due to additional scalar fields and charges, impacting both cosmological models and black hole shadows~\cite{Filho:2024ilq,Sekhmani:2024xyd,Gonzalez:2023rsd,Jusufi:2023xoa}. Additionally, quantum corrections from string T-duality provide alternative methods to avoid singularities, leading to stable black hole solutions consistent with quantum gravity expectations~\cite{Nicolini:2019irw}.

\begin{figure}
    \centering
\includegraphics[width=1\linewidth]{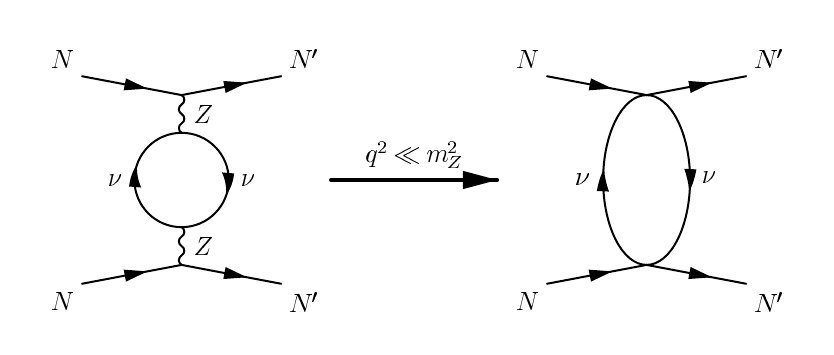}
    \caption{\textcolor{black}{Feynman diagrams in the Standard Model responsible for generating the Feinberg-Sucher long-range force, mediated by the exchange of two neutrinos within the four-Fermi effective theory \cite{Ferrer:1998rw}.}}
    \label{fig:Feynman}
\end{figure}

Most theoretical and experimental studies investigating long-range interactions have historically centered on electromagnetic and gravitational forces. However, seminal works such as the Casimir-Polder interaction \cite{casimir}, the Feinberg-Sucher force mediated by neutrino exchanges \cite{Feinberg:1968zz}, 
 as shown Feynman diagram in \ref{fig:Feynman} and recent advances involving supersymmetry and superstring theories \cite{Antoniadis:1997zg,Antoniadis:2000vd,Antoniadis:1986rn,Dimopoulos:1996vz}, illustrate ongoing interest in exotic long-range phenomena. Recent works have explored the possibility of detecting new forces and constraints from a variety of perspectives in both theoretical and experimental physics \cite{Chiao:2024hfu,Chiao:2023ezj}. For instance, superstring theories suggest that additional forces may become apparent at millimeter ranges \cite{Antoniadis:1997zg}, while alternative mechanisms mediated by Higgs and Goldstone bosons have also been proposed \cite{Ferrer:1998rw}. Similarly, extra dimensions and related string physics scenarios provide avenues for future collider experiments \cite{Antoniadis:2000vd} and have been discussed in the context of four-dimensional superstring models \cite{Antoniadis:1986rn}. On the experimental front, signatures of low-energy gauge mediated supersymmetry breaking are expected to produce unique signals \cite{Dimopoulos:1996vz}, and macroscopic forces originating from supersymmetry have been investigated in detail \cite{Dimopoulos:1996kp}. Recent precision tests of gravity, including measurements of the Casimir force, have been pivotal in constraining theories beyond the Standard Model \cite{Klimchitskaya:2023niz, Klimchitskaya:2020cnr, Tan:2020vpf}, while combined tests of the gravitational inverse-square law have further limited these possibilities \cite{Ke:2021jtj}. Moreover, advances in detector technologies have enabled new constraints on dark matter candidates, such as dark photons \cite{Chiles:2021gxk}. Finally, broader implications for fundamental physics, including space tests of the equivalence principle \cite{Battelier:2019kmc} and studies of CP and flavor violations in supersymmetric grand unified theories with right-handed neutrinos \cite{Hirao:2021pmh}, have been discussed in recent literature.

Experimental and theoretical studies have contributed significantly to our understanding of long-range forces in various physical contexts. For instance, Sukenik et al. \cite{Sukenik:1993zz} provided an early experimental measurement of the Casimir-Polder force, which has since been a reference point in precision tests of quantum electrodynamics. In parallel, theoretical investigations have explored how neutrino backgrounds and finite temperature effects can induce long-range forces \cite{Ferrer:1998ju, Ferrer:1999ad}. Recent studies have extended this line of inquiry, analyzing spin-dependent and parity-violating effects in neutrino-mediated forces \cite{Ghosh:2024qai}, as well as considering the impact of dark relics on background-induced forces \cite{Barbosa:2024pkl}. Additional work has addressed the detailed behavior of neutrino-mediated potentials in cosmological settings \cite{Blas:2022ovz} and has proposed novel experimental \textcolor{black}{approaches such} as probing atomic parity \textcolor{black}{violation to} detect the two-neutrino exchange force \cite{Ghosh:2019dmi}. Moreover, the interplay between Bose-Einstein condensation, dark matter, and acoustic peaks has been discussed as a mechanism potentially influencing these long-range interactions \cite{Ferrer:2004xj}. The theoretical framework for long-range forces has been further elaborated using quantum field theory at both zero and finite temperature \cite{Nowakowski:2000zd}. Finally, the influence of the cosmic photon heat bath on the Casimir-Polder force has also been studied \cite{Ferrer:1999tn}, and the possibility of mediating forces via Higgs and Goldstone bosons has been revisited in the context of fundamental interactions \cite{Ferrer:1998rw}. Recent advances in neutrino and gravitational physics have spurred a range of investigations into phenomena spanning from gravitational collapse to cosmic neutrino backgrounds. For example, Grifols et al. \cite{Grifols:1997jd} studied gravitino production in gravitational collapse scenarios, while Horowitz and Pantaleone \cite{Horowitz:1993kw} examined long-range forces induced by the cosmological neutrino background. Complementing these early works, Wilczek \cite{Wilczek:1982rv} discussed mechanisms of axion and family symmetry breaking that bear on dark matter searches.

More recently, efforts have been directed toward understanding the diffuse supernova neutrino background and its implications for new physics \cite{MacDonald:2024vtw} as well as exploring beyond-standard model effects in neutrino oscillations \cite{Calatayud-Cadenillas:2024wdw}. Investigations into the neutrino lifetime in both solar and core-collapse supernova contexts have further enriched our knowledge \cite{Martinez-Mirave:2024hfd}. Additionally, the synergy between cosmological and laboratory searches in neutrino physics has been emphasized to tighten constraints on these elusive particles \cite{Gerbino:2022nvz}.

The impact of the cosmic neutrino background extends even to gravitational phenomena, with studies examining its role in black hole superradiance \cite{Lambiase:2025twn} and setting limits on dark matter and ultralight scalars using precision techniques such as gyroscope spin measurements and atomic clocks \cite{Aliberti:2024udm}. Furthermore, quantum coherence effects in neutrino spin-flavor oscillations have been analyzed in detail \cite{Alok:2024xeg}, while astrophysical neutrino oscillations observed through pulsar timing array analyses offer fresh insights into neutrino behavior over cosmic distances \cite{Lambiase:2023pxd}. The interplay between gravitational waves and neutrino oscillations within the framework of Chern-Simons axion gravity has been explored to understand potential signatures of modified gravity \cite{Lambiase:2022ucu}, and extended theories of gravity have been tested using gamma-ray bursts as observational probes \cite{Mastrototaro:2021xcl}. These theoretical considerations extend further into neutrino-mediated interactions, incorporating both Dirac and Majorana neutrinos, and exploring thermal corrections related to relic cosmic neutrinos.

The main aim of this paper is to derive corrections to the Schwarzschild black hole metric arising from exotic long-range forces within the framework of quantum field theory. Specifically, we consider two distinct models characterized by potentials mediated by very light pseudoscalar bosons and neutrino exchange. We then systematically analyze how these long-range interactions influence observable astrophysical properties, such as the photon sphere and shadow radius. By explicitly extending the Schwarzschild geometry with additional pseudoscalar- and neutrino-mediated potential terms, we derive both analytical and numerical results using three complementary methods, providing clear predictions that can be tested through black hole imaging observations.

\section{Long Range Forces and Newtonian Potentials}

In this section, we discuss the Newtonian potentials within the framework of quantum field theory (QFT), focusing on two central issues: first, the emergence of long-range forces mediated by very light or massless particles, and second, the temperature dependence of such forces when the exchanged quanta are present in a thermal background. A notable example is the cosmic relic neutrino background, whose existence profoundly alters the asymptotic behavior of the two-neutrino exchange Feinberg-Sucher force.

The concept of force, originally introduced by Newton, continues to play a fundamental role in physics despite the advent of quantum theory and relativistic frameworks. Modern interaction theories frequently employ QFT, which elegantly bridges classical mechanics and quantum physics. This connection is particularly evident when QFT yields classical long-range forces and their quantum corrections. Specifically, quantum corrections to classical potentials such as electromagnetism and gravity have been thoroughly investigated. For example, QFT provides corrections to the classical Coulomb potential as  \cite{Nowakowski:2000zd}
\begin{align}
V_{\text{em}}(r)&=\frac{e^2}{r}\left[1 + \delta V^{\text{QM}}_{\text{em}}(r)\right],\\[6pt]
\delta V^{\text{QM}}_{\text{em}}(r)&=\frac{2\bar{\alpha}_{fine}}{3\pi}\left(\ln(1/m_er)-C -\frac{5}{6}\right)-\frac{2\bar{\alpha}_{fine}^2 e^2}{225\pi}\frac{1}{(m_er)^4},
\end{align}
\textcolor{black}{where $\bar{\alpha}_{fine}$ is the fine structure constant, $e$ is the charge of the electron, $m_e$ is the electron mass and $C=0.577$.} Similarly, quantum corrections to gravitational potentials have also been derived \cite{Donoghue:1993eb,Donoghue:1994dn,Burgess:2003jk}:
\begin{align}
V_{\text{gravity}}(r)&=-\frac{G_NM_1M_2}{r}\left[1 + \delta V^{\text{QM}}_{\text{gravity}}(r)\right],\\[6pt]
\delta V^{\text{QM}}_{\text{gravity}}(r)&=-\frac{G_N(M_1+M_2)}{r}+\frac{127G_N}{30\pi^2 r^2},
\end{align}
\textcolor{black}{where $G_N$ is the Newton’s gravitational constant.}

New long-range forces may arise naturally within the particle spectrum already established by experiments, mediated by neutrinos due to their negligible masses. Feinberg and Sucher provided pioneering calculations for these neutrino-mediated potentials, distinguishing between Dirac and Majorana neutrinos \cite{Feinberg:1968zz,Feinberg:1989ps,Ferrer:1998rw,Ferrer:1999tn,Ferrer:1999ad,Ferrer:2004xj}:
\begin{align}
V_{\text{Dirac}}(r)&=\frac{G_F^2m_{\nu}^3g_{V}g_{V}'}{4\pi^3r^2}K_3(2m_{\nu}r),\\[6pt]
V_{\text{Majorana}}(r)&=\frac{G_{F}^2m_{\nu}^2g_{V}g_{V}'}{2\pi^3r^3}K_{2}(2m_{\nu}r),
\end{align}
\textcolor{black}{where $G_F$ is the Fermi constant, $m_\nu$ is the neutrino mass, $g_V$, $g_{V}'$ denote vector coupling constants and $K_n$ are modified Bessel functions.} For massless neutrinos, these potentials reduce to the well-known Feinberg-Sucher form \cite{Feinberg:1968zz}:
\begin{equation}
V_{\text{FS}}(r)=\frac{G_{F}^2g_{V}g_{V}'}{4\pi^3r^5}.
\end{equation}
Theoretical extensions beyond the Standard Model suggest additional long-range forces, often mediated by exotic particles such as axions, majorons, scalar and pseudoscalar bosons appearing in supergravity theories. A compelling possibility includes towers of massive gravitons resulting from higher-dimensional compactifications. Experimental searches for such exotic forces continue actively, as their detection could offer profound insights into new physics.

A distinct and intriguing aspect of QFT is the temperature dependence of long-range forces when exchanged quanta are immersed in a thermal bath. Cosmic relic photons and neutrinos serve as practical examples. Within finite-temperature QFT frameworks, the propagators for fermions and bosons in a thermal bath have well-known structures, explicitly temperature-dependent through distribution functions  
\cite{Horowitz:1993kw,Ferrer:1998ju,Ferrer:1998rw,Ferrer:1999ad,Ferrer:1999tn,Ferrer:2004xj,Nowakowski:2000zd}. For calculating potentials from amplitudes, two primary methodologies exist. The first, widely used method involves Fourier-transforming static limit amplitudes:
\begin{equation}
V(r)=\frac{1}{2\pi^2 r}\int^{\infty}_0 dQ \, Q \, {\cal M}(Q)\sin(Qr),
\end{equation}
\textcolor{black}{where $Q$ denotes a generic U(1) charge.}
The alternative approach employs dispersion relations, defining potentials through discontinuities of the Feynman amplitudes \cite{Feinberg:1989ps}
\begin{equation}
V(r)=\frac{-i}{8\pi^2 r}\int^{\infty}_{4m^2} dt [{\cal M}]_t\exp(-\sqrt{t}r).
\end{equation}

In the following sections, we use these specific potential models to derive weak-force corrections to black hole solutions via long-range potentials.

\section{Weak-Force Corrections to Black Hole Solutions via Long-Range Potentials}

In this section, we analytically solve Einstein's gravitational equations incorporating Newtonian-type potentials arising specifically from the Feinberg–Sucher potential, mediated by massless neutrinos, and the Ferrer–Nowakowski potential, associated with massless pseudoscalar bosons. Such potentials are particularly relevant in scenarios involving weakly interacting fields, leading explicitly to corrections in the black hole metric characterized by weak coupling constants.
Considering a static and spherically symmetric spacetime, the general form of the line element can be written as:
\begin{equation}
ds^2
= -f(r)\,dt^2 +\frac{dr^2}{f(r)} +r^2(d\theta^2+\sin^2\theta\, d\phi^2).
\end{equation}
We assume the validity of Einstein's field equations, namely $G_{\mu\nu}=8\pi T_{\mu\nu}$, and investigate the corresponding gravitational equations sourced by these potentials. For the $tt$-component, the gravitational equation under consideration becomes:
\begin{equation}\label{grav_eq}
\frac{r f'(r)+f(r)-1}{r^2}=-8\pi \rho(r),
\end{equation}
where $\rho(r)$ denotes the energy density, directly related to the potential $V(r)$ through the Poisson-type equation \cite{Nicolini:2019irw,Filho:2024ilq}:
\begin{equation}\label{density_eq}
\rho(r)=\frac{1}{4\pi}\Delta V(r).
\end{equation}
The sign and form of the resulting potentials indicate that standard energy conditions may be violated in the interior region of these modified black hole solutions, reflecting nontrivial quantum field theoretical effects.
\subsection{Model 1:  Feinberg-Sucher potential for massless neutrino}
\begin{figure}
    \centering
\includegraphics[width=1\linewidth]{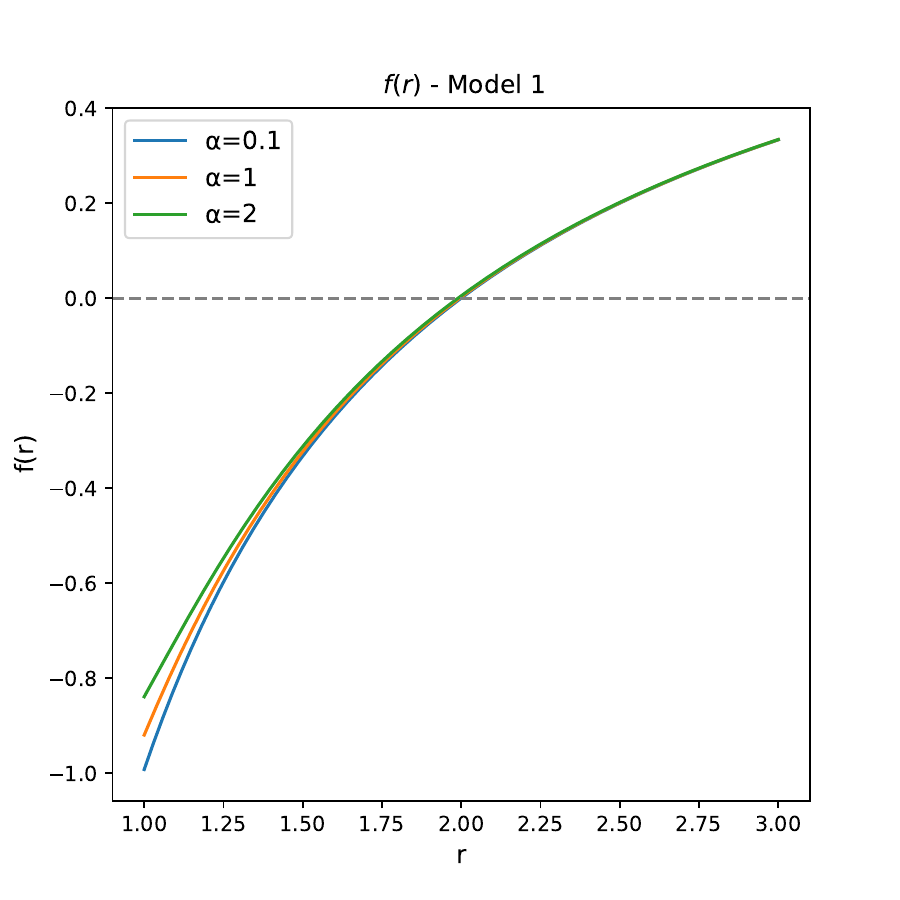}
    \caption{Figures shows the lapse function f(r) as a function of the radial distance r for \textbf{Model 1} ($\alpha=g_v g_{v'}$).}
    \label{fig:01}
\end{figure}

In this work, we use the Feinberg-Sucher potential for massless neutrinos, given by~\cite{Feinberg:1968zz}
\begin{equation}\label{FS}
V_{FS}(r) = \frac{G^2_F g_v g_{v'}}{4 \pi^3 r^5},
\end{equation}
\textcolor{black}{where $G_F$ is the Fermi constant and $g_v,g_{v'}$ are weak vector coupling constants.} It is worth noting that this Feinberg-Sucher potential is repulsive, due to an extra minus sign arising from the fermion loop contribution. Using equation (\ref{density_eq}) and the computed Laplacian, we get
\begin{equation}\label{density_final}
\rho(r)=\frac{1}{4\pi}\Delta V_{FS}(r)=\frac{5G^2_F g_v g_{v'}}{4\pi^4 r^7}.
\end{equation}
Substituting the obtained density (\ref{density_final}) into the gravitational equation (\ref{grav_eq}), we have
\begin{equation}
r\,f'(r)+f(r)-1+\frac{10\,G_F^2\,g_v\,g_{v'}}{\pi^3\,r^5}=0.
\end{equation}
Performing the integral, we obtain the solution for \\ \textbf{Model 1}: 
\begin{equation}\label{analytic_final}
f(r)=1-\frac{2M}{r}+\frac{5\,G_F^2\,g_v\,g_{v'}}{2\pi^3\,r^5}.
\end{equation}
Typically, the metric is asymptotically flat at large distances, $ \lim_{r\to\infty} f(r)=1.$
We have derived an explicit analytic solution for the metric function stemming from the Feinberg-Sucher potential, which is illustrated in Figure \ref{fig:01}.

\subsection{Model 2:  Ferrer-Nowakowski potential in boson-mediated long-range interactions at finite temperature}

We analytically derive the gravitational metric function arising from a finite-temperature corrected effective potential consisting of vacuum and thermal contributions. This potential is relevant in boson-mediated long-range interactions at finite temperature, contrasting with neutrino-mediated forces. The metric function is explicitly solved analytically, illustrating how finite-temperature corrections influence gravitational interactions mediated by very light bosons. We consider the total finite-temperature corrected potential given by~\cite{Ferrer:1998rw}:
\begin{equation}\label{tot}
V_{tot}(r)=V_T(r)+V(r) \simeq -\frac{3GG'}{16\pi^3 r^3},
\end{equation}
\textcolor{black}{where $G$ and $G'$ are the global coupling constants which capture the constants of the four vertices and the two Higgs propagators.} This potential combines vacuum and thermal contributions, providing a net attractive interaction, contrasting with neutrino-mediated interactions (the Feinberg-Sucher force is repulsive \ref{FS}), where finite-temperature corrections may reverse the sign of the force~\cite{Ferrer:1998ju}. Thus, the Laplacian explicitly is:
\begin{equation}\label{laplacian_final}
\Delta V_{tot}(r)=-\frac{9GG'}{8\pi^3 r^5}.
\end{equation} From (\ref{density_eq}), the density is obtained as:
\begin{equation}\label{density_final}
\rho(r)=\frac{1}{4\pi}\Delta V_{tot}(r)=-\frac{9GG'}{32\pi^4 r^5}.
\end{equation}
Substituting (\ref{density_final}) into (\ref{grav_eq}) yields:
\begin{equation}\label{final_ODE}
r\,f'(r)+f(r)-1-\frac{9GG'}{4\pi^3\,r^3}=0.
\end{equation}
The ODE (\ref{final_ODE}) is solved analytically using the integrating factor method. The explicit general solution for \textbf{Model 2} is thus:
\begin{equation}\label{analytic_final}
f(r)=1-\frac{2M}{r}-\frac{9GG'}{8\pi^3\,r^3}.
\end{equation}
The metric is asymptotically flat at large distances, $ \lim_{r\to\infty} f(r)=1.$
\begin{figure}
\includegraphics[width=1\linewidth]{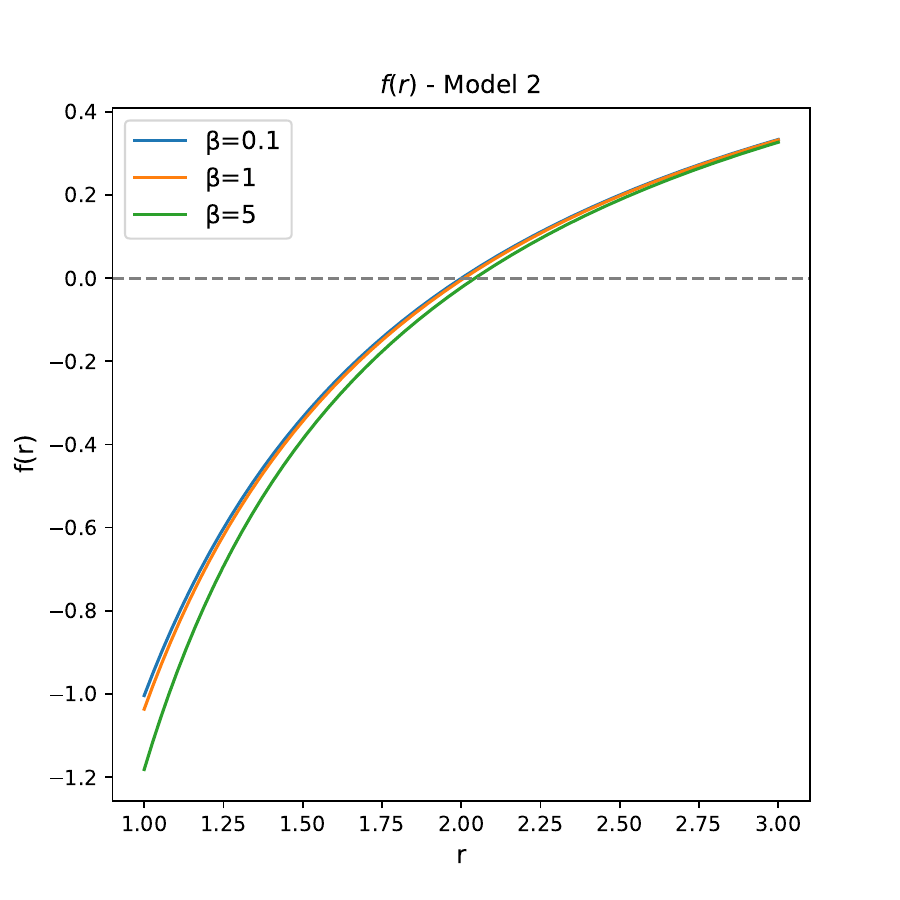}
    \caption{Figures shows the lapse function f(r) as a function of the radial distance r for \textbf{Model 2} ($\beta=GG'$ ).}
    \label{fig:02}
\end{figure}
We have obtained an explicit analytic solution of the metric function arising from the Ferrer-Nowakowski potential plotted in \ref{fig:02}.
Comparing the finite-temperature corrections of this boson-mediated interaction with those arising from neutrino-mediated interactions~\cite{Ferrer:1998ju}, we highlight a notable distinction: the boson-mediated force remains attractive even after including thermal effects. This difference arises due to the opposite signs of thermal corrections in the respective propagators positive for bosonic fields and negative for fermionic fields~\cite{Feinberg:1989ps}. Consequently, the intriguing phenomenon of sign reversal at finite temperature, previously reported for neutrino-mediated forces, does not manifest in the boson-mediated interactions considered here. The black hole mass is determined by solving the horizon condition, defined by setting the temporal metric component to zero, \( g_{00}(r=r_+) = 0 \). Specifically, this involves imposing
\begin{equation}
f(r_+) = 0,
\end{equation}
which directly relates the black hole mass \( M \) to the horizon radius \( r_+ \). For \textbf{Model 1}, characterized by coupling constants  \( g_v \), and \( g_{v'} \), the black hole mass explicitly reads
\begin{equation}
M(r_+) = \frac{r_+}{2}\left(1 + \frac{5G_F^2\,g_v\,g_{v'}}{2\pi^3\,r_+^5}\right),
\end{equation}
indicating a reduction from the Schwarzschild mass due to the presence of additional coupling parameters. For \textbf{Model 2}, defined by couplings \( G \) and \( G' \), the corresponding mass takes the form
\begin{equation}
M(r_+) = \frac{r_+}{2}\left(1 - \frac{9GG'}{8\pi^3\,r_+^3}\right),
\end{equation}
demonstrating an enhancement relative to the Schwarzschild case. Note that in the limit of vanishing coupling parameters and charges (\( gg_v'= GG' = 0 \)), both expressions reduce consistently to the Schwarzschild black hole mass:
\begin{equation}
M_+ = \frac{r_+}{2}.
\end{equation}
The Hawking temperature associated with the event horizon radius \( r_+ \) is obtained through the relation \cite{Akhmedov:2006pg}
\begin{equation}
T_+ = \frac{\kappa}{2\pi},
\end{equation}
where \( \kappa \) represents the surface gravity, defined as
\begin{equation}
\kappa = \left.\sqrt{-\frac{1}{2}\nabla_\mu \chi_\nu \nabla^\mu \chi^\nu}\right|_{r=r_+} = \frac{1}{2}\left.\frac{df(r)}{dr}\right|_{r=r_+},
\end{equation}
\textcolor{black}{where \(\chi^\mu\) is the timelike Killing vector field generating the horizon, \(\nabla_\mu\) denotes the covariant derivative.} This quantity measures the gravitational acceleration at the horizon, thus determining the temperature of Hawking radiation. Explicitly, the Hawking temperature for \textbf{Model 1} becomes
\begin{equation}
T(r_+) =  \frac{1}{4\pi r_+} - \frac{5 G_F^2 g_v g_{v'}}{2\pi^4 r_+^6},
\end{equation}
showing corrections arising from the coupling terms which modify both the magnitude and the scaling behavior with the horizon radius. Similarly, for \textbf{Model 2}, the Hawking temperature explicitly reads
\begin{equation}
T(r_+) = \frac{1}{4\pi r_+} + \frac{9GG'}{16\pi^4\,r_+^4},
\end{equation}
which again clearly demonstrates how the presence of coupling constants influences the thermodynamic properties of the black hole. In both scenarios, the Schwarzschild limit (\(g g_v, GG' \to 0 \)) reproduces the well-known Schwarzschild results:
\begin{equation}
T = \frac{1}{8\pi M}.
\end{equation}
These expressions highlight explicitly how modifications to gravitational interactions alter fundamental thermodynamic relations in non-standard gravitational theories.

\section{Shadow of the black hole}

\subsection{Vertogradov-\"Ovg\"un Method: Effect of parameters on the photon sphere and the shadow radius} 

Here, we use the method elaborated by Vertogradov and \"Ovg\"un in the paper~\cite{Vertogradov:2024qpf}, which states that one can consider the lapse function $F$ in the form \begin{equation}F(r)=\left(1-\frac{2M}{r}\right)e^{\gamma g(r)}. \label{ff} \end{equation} \textcolor{black}{ Here $M$ is the mass of a black hole and  and $\gamma g(r)$ is the minimal geometrical deformation of the Schwarzschild spacetime. }\textcolor{black}{Then, we consider the radius  of a photon sphere as $r_{ph}^{(0)}=3M$.} According to the results of \cite{Vertogradov:2024qpf}, the following theorems hold.

\textbf{ Theorem 1:} If the seed metric is the Schwarzschild one with $r_{ph}^{(0)}=3M$, then the additional matter field increases the radius of a photon sphere if $g'(r_{ph}^{(0)})>0$ and decreases it if $g'(r_{ph}^{(0)})<0$  \cite{Vertogradov:2024qpf} . 

\textbf{ Theorem 2:} We can observe that if $g\left(r_{ph}^{(0)}\right)>0$ , then the shadow size decreases due to the additional matter field. Conversely, if 
$g\left(r_{ph}^{(0)}\right)<0$, then the shadow increases where $r_{ph}^{(0)}=3M$. \cite{Vertogradov:2024qpf} . 

Here, we assume small deviations from Schwarzschild spacetime and radius of a photon sphere decreases and increases in comparison with Schwarzschild case depending on the sign of $r_1$. In order to proceed we use method elaborated in the paper~\cite{Vertogradov:2024qpf}. Use the Eq. \ref{ff} and find the relation $\gamma$ for the \textbf{Model 1} and the \textbf{Model 2}, by solving  $F(r)=f(r)$ for $g(r)$ to check photon sphere and shadow radius behaviors.

For \textbf{Model 1}:
\begin{equation}
\left(1-\frac{2M}{r}\right)e^{\gamma g(r)}=1-\frac{2M}{r}+\frac{5G^2_F \alpha}{2\pi^3 r^5},     
\end{equation}
and solve it for $g(r)$ gives:
\begin{equation}
g(r)=  \frac{\log \left(1-\frac{5 \alpha  G_F^2}{2 \pi ^3 r^4 (2 M-r)}\right)}{\gamma },
\end{equation}
and
\begin{equation}
g'(r)=
\frac{5 \alpha  G_F^2 (8 M-5 r)}{\gamma  r (2 M-r) \left(2 \pi ^3 r^4 (2 M-r)-5 \alpha  G_F^2\right)},\end{equation} derivatives are taken with respect to $r$, indicated by the prime notation.

For \textbf{Model 2}:

\begin{equation}
\left(1-\frac{2M}{r}\right)e^{\gamma g(r)}=1-\frac{2M}{r}-\frac{9 \beta}{8\pi^3 r^3},     
\end{equation}
and solve it for $g(r)$ gives:
\begin{equation}
 g(r)=  \frac{\log \left(\frac{9 \beta }{8 \pi ^3 r^2 (2 M-r)}+1\right)}{\gamma },
\end{equation}
and 
\begin{equation}
g'(r)=
\frac{9 \beta  (3 r-4 M)}{\gamma  r (2 M-r) \left(9 \beta +8 \pi ^3 r^2 (2 M-r)\right)}.
\end{equation}

\begin{figure}
    \centering
\includegraphics[width=1\linewidth]{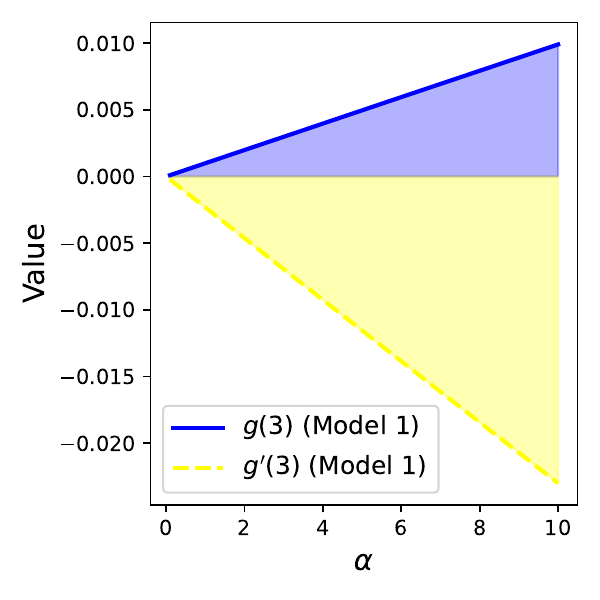}
\includegraphics[width=1\linewidth]{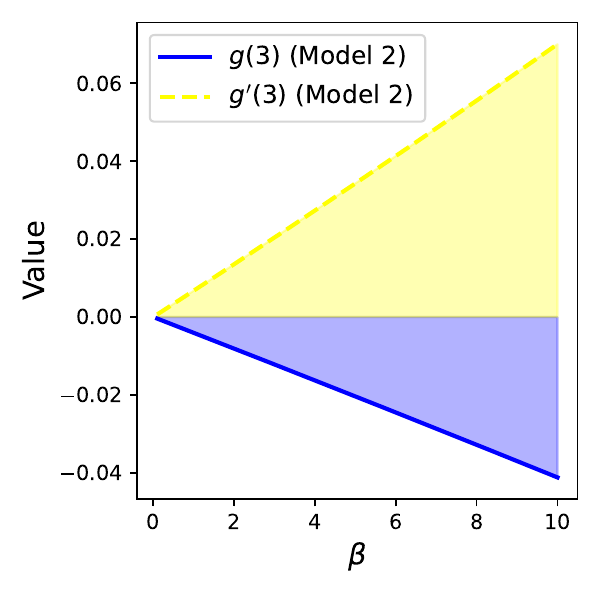}
    \caption{Figures shows the behavior of \(g(3)\) and \(g'(3)\) for Models 1 and 2, respectively.}
    \label{fig:2}
\end{figure}

\textcolor{black}{Note that we define as $\alpha=g_v g_{v'}$ for the weak vector coupling constants $g_vg_{v'}$ and $\beta=GG'$ for the global coupling constants $GG'$.} We now analyze the behavior of the effective function \(g(r)\) and its radial derivative \(g'(r)\) evaluated at the Schwarzschild photon–sphere \(r_{\mathrm{ph}}^{(0)}=3M\) (with \(M=1\)) for two distinct models. In both cases, the effective metric is parametrized by a function \(g(r)\) that encodes deviations from the Schwarzschild geometry due to additional matter fields. For simplicity, we set the constants \(G_F = 1\), and \(M = 1\). The coupling constants are denoted by \(\alpha\) (for \textbf{Model 1}) and \(\beta\) (for \textbf{Model 2}) to plot $g(r)$ and $g'(r)$. Figures \ref{fig:2} show the behavior of \(g(3)\) and \(g'(3)\) for Models 1 and 2 at photon sphere of the seed metric $r_{ph}=3$, respectively, as functions of the coupling parameters \(\alpha\) and \(\beta\) over the range \(0\le\alpha,\beta\le 10\).
For \textbf{Model 1} we have \(g(3)>0\) and \(g'(3)<0\), indicating that the photon sphere shrinks and the shadow becomes smaller due to the repulsive potential of massless neutrinos. In contrast, for \textbf{Model 2}, we note that \(g(3)<0\) for any \(\alpha>0\) (since the argument of the logarithm is less than unity), while \(g'(3)>0\) for small positive \(\alpha\). According to Theorem 1, this implies that the photon sphere radius increases with the potential of attractive interaction caused by boson-mediated long range interactions at finite temperature with vacuum and thermal contributions, and by Theorem 2 the shadow size is enlarged relative to the Schwarzschild geometry.  These results corroborate the theoretical predictions and provide clear visual confirmation of how the sign of \(g(r_{ph}^{(0)})\) and its derivative influence the observables in black hole imaging. This method is particularly valuable when an analytical derivation of the photon sphere and shadow radius is not feasible. It enables a simple investigation of how these quantities behave under various parameter settings.

\subsection{Vertogradov-\"Ovg\"un-Kobialko-Gal’tsov Method: Analytic approximations for photon sphere and shadow radius }

In this section, we derive the photon sphere radius and the squared black hole shadow radius for two distinct black hole models, utilizing analytic expansions around the black hole solutions in Feinberg-Sucher and Ferrer-Nowakowski potentials, using the method initially introduced by Vertogradov and \"Ovg\"un in \cite{Vertogradov:2024dpa} and subsequently improved by Kobialko and Gal’tsov in \cite{Kobialko:2024zhc}. We consider metrics of the form 
\begin{equation}
f(r)=1-\frac{2M}{r}+\sum_{i=2}^{n}\frac{c_i}{r^i},
\end{equation}
where $M$ is the black hole mass and $c_i$ represent coupling parameters arising from additional matter fields. Using expansions around the Schwarzschild photon sphere radius $r_{\mathrm{ph}}^{(0)}=3M$, the photon sphere radius and shadow radius squared are approximated as follows:
\begin{align}
r_{\mathrm{ph}}&=3M-\frac{1}{2}\sum_{i=2}^{n}\frac{(i+2)\,c_i}{(3M)^{i-1}}+\mathcal{O}(c_i),\\[6pt]
R_{\mathrm{sh}}^{2}&=27M^2-81M^2\sum_{i=2}^{n}\frac{c_i}{(3M)^i}+\mathcal{O}(c_i).
\end{align}

\subsubsection{Model 1}

The first model considered has the metric function
\begin{equation}
f(r)=1-\frac{2M}{r}+\frac{5G_F^2\alpha}{2\pi^3 r^5},
\end{equation}
which corresponds to a single additional term with $n=5$, identifying the coupling parameter as
\begin{equation}
c_5=\frac{5G_F^2\alpha}{2\pi^3},\quad c_i=0\text{ for }i\neq 5.
\end{equation}
The photon sphere radius and shadow radius squared for this model thus become
\begin{align}
r_{\mathrm{ph}}&=3M-\frac{35 G_F^2\alpha}{324\,\pi^3M^4},\\
R_{\mathrm{sh}}^{2}&=27M^2-\frac{5G_F^2\alpha}{6\pi^3M^3}.
\end{align}

\subsubsection{Model 2}

\begin{figure}
    \centering
\includegraphics[width=1\linewidth]{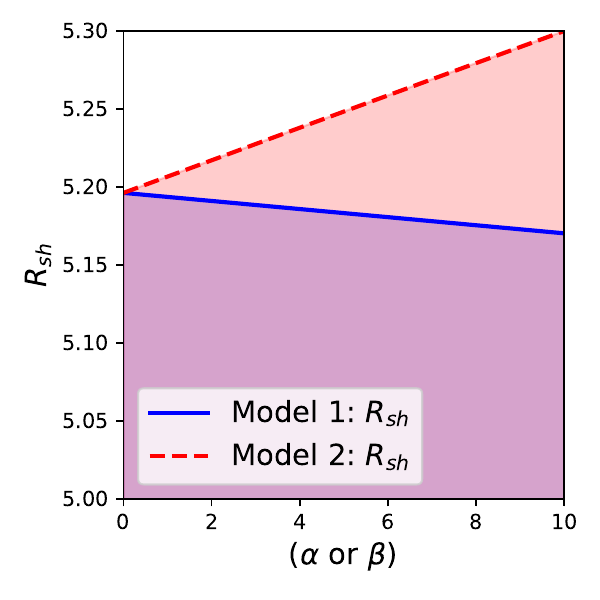}
    \caption{Figure shows shadow radius $R_{sh}$ versus parameters $\alpha$ and $\beta$}
    \label{fig:shadow3}
\end{figure}

The second model studied is given by
\begin{equation}
f(r)=1-\frac{2M}{r}-\frac{9\beta}{8\pi^3 r^3},
\end{equation}
identifying a single coupling parameter with $n=3$:
\begin{equation}
c_3=-\frac{9\beta}{8\pi^3},\quad c_i=0\text{ for }i\neq 3.
\end{equation}
The corresponding photon sphere radius and shadow radius squared yield
\begin{align}
r_{\mathrm{ph}}&=3M+\frac{5\beta}{16\,\pi^3\,M^2},\\
R_{\mathrm{sh}}^{2}&=27M^2+\frac{27\beta}{8\,\pi^3\,M}.
\end{align}
These results reveal distinctly different behaviors induced by the respective matter fields in the two models as shown in \ref{fig:shadow3}: \textbf{Model 1} exhibits a decrease in both the photon sphere radius and shadow radius for positive coupling $\alpha$. The shadow and photon sphere thus contract relative to their Schwarzschild counterparts, suggesting weaker gravitational lensing. In contrast, \textbf{Model 2:} both the photon sphere radius and shadow size are increased by the positive coupling $\beta$. These analytic expansions provide clear insights into how different coupling terms and radial dependencies affect crucial observational signatures, such as the size of black hole shadows and photon spheres. 
This method proves valuable in scenarios where analytical determination of the photon sphere and shadow radius is not feasible.   

\subsection{Photon sphere and black hole shadow: Null geodesics approach}

To verify and complement the analytic results obtained in the previous sections, we employ the standard null geodesics method, widely used in the literature to study black hole shadows and photon spheres (see, for example, Refs.~\cite{Falcke:1999pj,Cunha:2018acu,Atamurotov:2013sca,Cunha:2016wzk,Vagnozzi:2019apd,Shaikh:2018lcc,Cunha:2017eoe,Tsukamoto:2014tja,Perlick:2015vta,Bisnovatyi-Kogan:2010flt} and references therein). Investigating the black hole shadow geometry has recently attracted significant attention due to its observational relevance in the context of the Event Horizon Telescope (EHT) and its potential to constrain new physics beyond general relativity. The dynamics of photon trajectories in the equatorial plane (\(\theta=\pi/2\)) can be effectively studied using the Hamilton–Jacobi formalism, starting from the Hamiltonian for null geodesics given by:
\begin{equation}\label{hj}
H=\frac{1}{2} g^{\mu\nu} p_{\mu} p_{\nu}=\frac{1}{2}\left(\frac{L^{2}}{r^{2}}-\frac{E^{2}}{f(r)}+\frac{\dot{r}^{2}}{f(r)}\right)=0,
\end{equation}
where \( p_{\mu} \) denotes the photon four-momentum, \( \dot{r}=\partial H/\partial p_r \), \( E = -p_t \) is the photon energy, and \( L = p_\phi \) is the angular momentum. From Eq.~\eqref{hj}, the effective radial potential for photon motion can be expressed as:
\begin{equation}\label{effpot}
V(r)=f(r)\left(\frac{L^{2}}{r^{2}}-\frac{E^{2}}{f(r)}\right),
\end{equation}
with the null geodesic equation simplified to \( V(r) + \dot{r}^{2}=0 \).
The photon sphere corresponds to unstable circular orbits of photons, which satisfy the conditions  \cite{Claudel:2000yi}
\begin{equation}\label{unstableconditions}
V(r_p)=0,\quad V'(r_p)=0,\quad V''(r_p)<0.
\end{equation}
By introducing the impact parameter \( b=L/E=r_p/\sqrt{f(r_p)} \), the radius of the photon sphere \( r_p \) can be determined by solving the characteristic photon-sphere equation \cite{Claudel:2000yi}:
\begin{equation}\label{phsp}
\frac{f'(r_p)}{f(r_p)}=\frac{2}{r_p}.
\end{equation}
Due to the complexity of Eq.~\eqref{phsp}, analytic solutions for \( r_p \) typically are not possible for metrics with corrections beyond Schwarzschild geometry. Therefore, numerical methods are employed to find the largest real root, corresponding physically to the photon sphere radius. Once \( r_p \) is numerically determined, the black hole shadow radius \( R_s \), as observed by a static observer located at \( r_0 \), is given by:
\begin{equation}\label{shadowradius}
R_{s}=r_{p}\sqrt{\frac{f\left(r_{0}\right)}{f\left(r_{p}\right)}}.
\end{equation}
For an observer situated at spatial infinity (\(r_0\to\infty\)), the metric function satisfies \( f(r_0)\to 1 \), simplifying Eq.~\eqref{shadowradius} to:
\begin{equation}\label{shadowradiusinfty}
R_{s}= \frac{r_{p}}{\sqrt{f(r_{p})}}.
\end{equation}
To illustrate our results clearly and facilitate direct comparison, we adopt simplified notation by defining the dimensionless coupling parameters \(\alpha=g_v g_{v'}\) (for \textbf{Model 1}) and \(\beta=GG'\) (for \textbf{Model 2}).

\begin{equation} \label{shadow}
R_{s}^2=\frac{ r_p^{2}}{A(r_p)}.
\end{equation}
\begin{table}[htbp]
  \centering
  \caption{For \textbf{Model 1}: Photon radius and Shadow radius with various $\alpha$}
  \label{tab:dataset}
  \begin{tabular}{|c|c|c|}
    \hline
    $\alpha$ & $r_{\text{ph}}$ & $R_{\text{sh}}$ \\ \hline
0.1 & 2.99965 & 5.19589 \\
\hline
0.6 & 2.9979  & 5.1946 \\
\hline
1.1 & 2.99615 & 5.1933 \\
\hline
1.6 & 2.99438 & 5.19199 \\
\hline
2.1 & 2.99261 & 5.19068 \\ \hline
  \end{tabular}
\end{table}

\begin{table}[htbp]
  \centering
  \caption{For \textbf{Model 2}: Photon radius and Shadow radius with various $\alpha$}
  \label{tab:dataset2}
  \begin{tabular}{|c|c|c|}
    \hline
    $\beta$ & $r_{\text{ph}}$ & $R_{\text{sh}}$ \\ \hline
0.1 & 3.00101 & 5.1972 \\
\hline
0.6 & 3.00602 & 5.20242 \\
\hline
1.1 & 3.01101 & 5.20761 \\
\hline
1.6 & 3.01596 & 5.21277 \\
\hline
2.1 & 3.02087 & 5.2179 \\
\hline
  \end{tabular}
\end{table}
Our numerical results, summarized in Table I, indicate a clear trend: for increasing values of the coupling parameters, the photon sphere radius and corresponding shadow size exhibit distinct behavior for each model. Specifically, the presence of a repulsive potential term (\textbf{Model 1}) leads to a decrease in photon sphere radius and shadow size, signaling gravitational screening effects. In contrast, the attractive potential term (\textbf{Model 2}) yields an increase in these quantities, reflecting stronger gravitational attraction relative to the Schwarzschild case.  These numerical outcomes are fully consistent with the analytical findings obtained from the previous theoretical approaches, highlighting the robustness and reliability of our analytic expansions.

\subsection{Analytical quasinormal modes in the eikonal regime}

Recent studies on black hole dynamics and gravitational wave signatures have underscored the significance of analyzing perturbations of black hole spacetimes. The seminal works of Press \cite{Press:1971wr} and Goebel \cite{1972ApJ...172L..95G} introduced foundational concepts regarding gravitational wave emissions from perturbed black holes which has been demonstrated that the so-called “vibrations of a black hole,” originally described by Press \cite{Press:1971wr}, correspond to gravitational waves trapped in spiral trajectories near the well-known unstable photon orbit at \( r = 3M \) in the Schwarzschild spacetime. Similar trapped modes, or vibrations, have also been analyzed for spinning (Kerr) black holes. Importantly, these gravitational wave vibrations do not represent permanent sources of radiation; rather, they temporarily store gravitational wave energy at high frequencies before it escapes or is absorbed by the black hole \cite{1972ApJ...172L..95G}. Such perturbations, especially in the context of scalar fields around Schwarzschild black holes, have been systematically explored using phase integral methods \cite{Andersson:1995vi}. Further insights into black hole perturbations and their observational signatures, particularly related to photon spheres and shadow observations, have been extensively discussed in the literature. Cardoso et al. investigated the stability and observational properties of photon spheres, emphasizing their role in defining quasinormal modes and observational features such as black hole shadows \cite{Cardoso:2008bp, Cardoso:2014sna}. If the metric is asymptotically flat, the quantity \( b_{\text{critical}} \) defines the critical impact parameter below which photons cannot escape back to infinity and instead fall into the black hole. Perturbation modes of fields propagating around black holes may become trapped in the vicinity of photon spheres, giving rise to long-lived quasi-normal modes (QNMs) \cite{Cardoso:2014sna}. Following the analysis presented in \cite{Cardoso:2008bp}, one finds that in the large angular momentum limit (\( l \rightarrow \infty \)), the real part of the QNM frequencies can be approximated as follows. Moreover, photon spheres have been studied beyond General Relativity within frameworks including supergravity, demonstrating that distinct photon sphere characteristics can serve as observational signatures for alternative theories of gravity \cite{Cvetic:2016bxi}. Analytical methods, notably in the eikonal approximation, have provided powerful tools to understand quasinormal modes around various black hole spacetimes, offering direct comparisons to observational data \cite{Churilova:2019jqx}. Such analytical results are particularly relevant for interpreting deviations from standard black hole geometries arising from exotic long-range interactions, which we explicitly investigate in the present work.

Now we calculate the quasinormal modes (QNMs) for two modified Schwarzschild metrics arising from long-range forces mediated by pseudoscalar bosons. The QNM frequencies are obtained using an analytic approximation method in the eikonal limit, following the general WKB-based expansion approach for asymptotically flat metrics \cite{Churilova:2019jqx}:

\begin{equation}\label{expMetric}
f(r)=1-\frac{2M}{r}+\frac{\lambda_2}{r^2}+\frac{\lambda_3}{r^3}+\frac{\lambda_4}{r^4}+\frac{\lambda_5}{r^5}+ {\cal O}\left(\frac{1}{r^6}\right),
\end{equation}
with \textcolor{black}{coefficients \(\lambda_i\) characterizing deviations from the Schwarzschild solution}. Correspondingly, the QNM frequencies $\omega$ can be expressed analytically in the eikonal limit as \cite{Churilova:2019jqx}:
\begin{widetext}
\begin{eqnarray}\label{QNMGen}
\omega=\frac{\left(\ell+\frac{1}{2}\right)}{3\sqrt{3}M}\left(1+\frac{\lambda_2}{6 M^2}+\frac{\lambda_3}{18 M^3}+\frac{\lambda_4}{54 M^4}+\frac{\lambda_5}{162 M^5}\right)-i\,\frac{\left(n+\frac{1}{2}\right)}{3\sqrt{3}M}\left(1+\frac{\lambda_2}{18 M^2}-\frac{\lambda_3}{27 M^3}-\frac{\lambda_4}{27 M^4}-\frac{11 \lambda_5}{486 M^5}\right)\\+
{\cal O}\left(\frac{1}{\ell +\frac{1}{2}}\right) \notag,
\end{eqnarray}
\end{widetext}
where \(\ell\) and \(n\) denote the angular and overtone numbers, respectively. We apply this general formalism to two concrete examples motivated by quantum field theoretic scenarios of long-range forces mediated by very light pseudoscalar bosons.

\subsection{Model 1}

In the first scenario, the metric function has the form:
\begin{equation}\label{Model1}
f(r)=1-\frac{2M}{r}+\frac{5G_F^2\alpha}{2\pi^3 r^5}.
\end{equation}
Matching this metric function to the general expansion (\ref{expMetric}), we clearly identify the non-zero coefficient:
\begin{equation}
\lambda_5=\frac{5G_F^2\alpha}{2\pi^3},\quad \lambda_2=\lambda_3=\lambda_4=0.
\end{equation}
Inserting these values into Eq.~(\ref{QNMGen}), we obtain the analytic expression for the QNM frequencies $\omega$ of Model~1:
\begin{eqnarray}\label{QNMmodel1}
\omega&=&\frac{\left(\ell+\frac{1}{2}\right)}{3\sqrt{3}M}\left(1+\frac{5G_F^2\alpha}{324\pi^3 M^5}\right)\nonumber\\
&-&i\,\frac{\left(n+\frac{1}{2}\right)}{3\sqrt{3}M}\left(1-\frac{55G_F^2\alpha}{972\pi^3 M^5}\right).
\end{eqnarray}
The real frequency is positively corrected, indicating slightly increased oscillation frequencies for positive \(\alpha\), whereas the imaginary part is less negative, again signaling a reduced damping rate. This means that in the repulsive potential of massless  neutrinos, perturbations around the black hole ring at higher frequencies and also last longer relative to the Schwarzschild solution.

\subsection{Model 2}

In the second scenario, the metric function has the form:
\begin{equation}\label{Model2}
f(r)=1-\frac{2M}{r}-\frac{9\beta}{8\pi^3 r^3}.
\end{equation}
Again, comparing with Eq.~(\ref{expMetric}), the non-zero coefficient is:
\begin{equation}
\lambda_3=-\frac{9\beta}{8\pi^3},\quad \lambda_2=\lambda_4=\lambda_5=0.
\end{equation}
Substituting this into Eq.~(\ref{QNMGen}) yields the quasinormal frequencies for Model~2:
\begin{eqnarray}\label{QNMmodel2}
\omega^{(2)}&=&\frac{\left(\ell+\frac{1}{2}\right)}{3\sqrt{3}M}\left(1-\frac{\beta}{16\pi^3 M^3}\right)\nonumber\\
&-&i\,\frac{\left(n+\frac{1}{2}\right)}{3\sqrt{3}M}\left(1+\frac{\beta}{24\pi^3 M^3}\right).
\end{eqnarray}
The real part of the frequency exhibits a negative correction relative to Schwarzschild, implying slightly lower oscillation frequencies for positive couplings. Conversely, the imaginary part receives a positive contribution, leading to smaller absolute values, thereby reducing the damping rate and increasing the lifetime of these QNMs. Physically, this suggests that the attractive potential caused by boson-mediated long range interactions at finite temperature tends to create a longer-lived gravitational perturbation around the black hole, enhancing observational prospects. The analytical results for the QNMs derived here clearly demonstrate that different quantum field-theoretic couplings lead to distinct observable effects on black hole oscillation spectra. In particular, the presence of higher-order corrections (\(r^{-3}\) or \(r^{-5}\)) impacts both oscillation frequencies and lifetimes of gravitational perturbations. The imaginary part modifications are of particular observational interest due to their influence on damping times. This influence suggests potential detectability of resulting gravitational waveform changes by future gravitational wave observatories such as LISA.

\section{Conclusions}
\label{conclusions}

In this work, we derived new black hole solutions which are modifications to the Schwarzschild black hole geometry arising from exotic long-range interactions predicted by quantum field theory. Specifically, we considered two distinct theoretical models characterized by  potentials mediated by very light pseudoscalar bosons and neutrino-pair exchanges. Employing both analytical expansions and numerical methods, we computed corrections to observable quantities such as the photon sphere radius, shadow size, and quasinormal mode frequencies. Our findings reveal that these long-range interactions produce distinct and measurable deviations from classical Schwarzschild predictions.

In \textbf{Model 1}, characterized by neutrino-mediated potentials, we observed a reduction of both photon sphere and shadow radii, consistent with theoretical expectations stemming from the repulsive potential of massless neutrinos. Conversely, \textbf{Model 2}, governed by the attractive potential caused by boson-mediated long range interactions at finite temperature with vacuum and thermal contributions, exhibited an enlargement of the photon sphere and shadow size. These contrasting behaviors highlight the sensitivity of black hole shadows to the underlying matter fields and interactions, providing potential observational signatures of physics beyond the Standard Model.

Furthermore, we applied and cross-validated our results through three complementary methods for calculating shadow radius. All methods consistently confirmed the impact of these exotic interactions, underscoring their robustness and physical relevance. The analytic approach presented here, which utilizes expansions of the lapse function and the corresponding corrections to the photon sphere and shadow radii, provides a transparent means to assess the impact of exotic long-range forces on black hole imaging. Our analysis indicates that observations of black hole shadows, such as those conducted by the Event Horizon Telescope, offer a promising avenue to detect subtle quantum field-theoretic corrections. Future precision measurements may thus provide unique tests of long-range forces predicted by extended particle physics models and significantly enhance our understanding of fundamental interactions in strong gravitational fields.

\acknowledgements
This work is dedicated to the memory of Durmus Demir, whose exceptional contributions to physics and unwavering passion will forever inspire generations to come. A.{\"O}. would like to acknowledge the contribution of the COST Action CA21106 - COSMIC WISPers in the Dark Universe: Theory, astrophysics and experiments (CosmicWISPers), the COST Action CA21136 - Addressing observational tensions in cosmology with systematics and fundamental physics (CosmoVerse), the COST Action CA22113 - Fundamental challenges in theoretical physics (THEORY-CHALLENGES), the COST Action CA23130 - Bridging high and low energies in search of quantum gravity (BridgeQG) and the COST Action CA23115 - Relativistic Quantum Information (RQI) funded by COST (European Cooperation in Science and
Technology). We also thank EMU, TUBITAK (Turkiye) and SCOAP3 (Switzerland) for their support. 

\bibliographystyle{apsrev4-1}

\bibliography{main.bib}

\end{document}